# Affine nonmagnetic transformation optics and its application to a practical bending adapter design

Hongyi Xu[1], Baile Zhang[1, 2, 3*], Tianyuan Yu[1], George Barbastathis[2, 3], Handong Sun[1*]

1. Division of Physics and Applied Physics, School of Physical and Mathematical Sciences, Nanyang Technological University, Singapore 637371, Singapore.
2. Singapore-MIT Alliance for Research and Technology (SMART) Centre, Singapore 117543, Singapore.
3. Department of Mechanical Engineering, Massachusetts Institute of Technology, Cambridge, Massachusetts 02139, USA.
*Corresponding authors, electronic mail: blzhang@ntu.edu.sg, hdsun@ntu.edu.sg

One of the bottlenecks that limit the transition of transformation-optics devices from concepts to practical use is the non-unit magnetic permeability generally required from a mathematical transformation. Simple renormalization of permeability, as used in many previous designs and experiments, introduces impedance mismatch and thus degrades the functional photonic performance. Here we propose an area-preserving affine coordinate transformation as a general method to solve this problem. Ideal transformation-optics functions can be preserved while nonmagnetism is achieved. As a specific example, we illustrate how to apply this affine method into the design of a two-dimensional electromagnetic beam bending adapter. Concerns related to fabrication, such as anisotropy degree and bending angles, are fully discussed. Our study is a significant step toward practical use of ideal transformation optics devices that can be implemented directly with existing dielectric materials.

Transformation optics, first proposed in the context of invisibility cloaking [1, 2], is also applicable to a broad variety of electromagnetic wave converters [3-5]. The original conception of implementing transformation optics necessitates the utilization of media that are both inhomogeneous and anisotropic. To push the implementation into the optical regime, inhomogeneity-only quasiconformal mapping technique was proposed [6], but the manufacturing cost remains high and undesired distortion inevitably appears in the output rays [7]. To address this challenge, homogeneous transformation methods [8, 9] were recently proposed and experimentally demonstrated [10-12]. Their utility became immediately clear after the first two attempts of invisibility cloak [10, 11] were realized at the macroscopic scale and visible wavelengths.

Despite of this success, a second limitation to date has still remained in the practical adoption of transformation optics devices: The transformation typically results in non-unit values for the magnetic permeability $\mu$, which is not known generally how to implement in optical wavelengths with conventional optical materials. Although conformal mapping can help to achieve nonmagnetism in some cases [2], its application introduces inhomogeneity which is hard to fabricate, and suffers from the limitation of fixed conformal module.

Two approaches have been proposed to tackle this challenge. The first approach, which is commonly used in the literature to date, is to renormalize both the permittivity and permeability such that the permeability is forced to be unity [13-15]. However, the renormalization introduces impedance mismatch at interfaces between regions with different transformation kernels. That is, the basic transformation optics premise of invariance in Maxwell's equations is violated, resulting in reflection and scattering. The second possible approach, which on the other hand has received much less attention, is to maintain the transformation Jacobian at the value of one throughout space. This was first suggested in the context of inhomogeneous transformation design [16, 17]. However, a general nonmagnetic transformation method toward the second approach has not been discussed thoroughly, nor have any experimental implementations been reported, to our knowledge.

Here we introduce a transformation optical design which possesses simultaneously a unitary (*i.e.* nonmagnetic*) Jacobian and piece-wise homogeneity (*i.e.*, it is implementable by homogeneous anisotropic materials or metamaterials). Our approach admits both macroscale and nanoscale fabrication. In the macroscopic regime, our design can be achieved using natural birefringence materials such as Calcite [10, 11]. Nanoscale realizations are available via subwavelength anisotropic patterning (*i.e.* form birefringence [18, 19]). The homogeneous implementation is much less challenging than the inhomogeneous case, due to the proximity effect correction (PEC) in electron beam lithography [20]. Moreover, the geometrical approach simplifies the problem to one of graphical design that in many cases can be carried out analytically.

Consider a general two-dimensional (2D) affine coordinate transformation $x' = ax + by + c$, $y' = dx + ey + f$, from triangle AOB to A'O'B' in the *x-y* plane, as shown in Fig. 1(a). The associated Jacobian matrix is $\bar{\bar{J}} = [a,b;d,e]$. For transverse magnetic (TM) modes where the H field is perpendicular to the *x-y* plane, we obtain the transformed relative dielectric permittivity and permeability as $\bar{\bar{\varepsilon}}' = \bar{\bar{J}}\bar{\bar{J}}^T / \det(\bar{\bar{J}})$, $\mu' = 1/\det(\bar{\bar{J}})$.

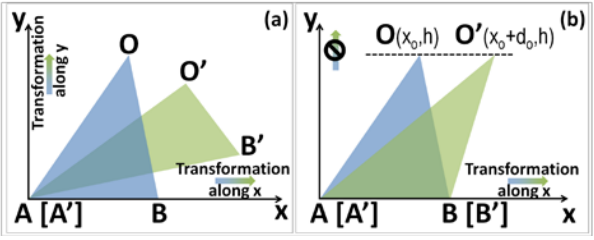

**Figure 1.** (Color online) (a) The area-preserving transformation from triangle *AOB* (blue) to *AO'B'* (green) along both *x* and *y* directions. (b) Special case of horizontal shear along the *x* direction only.

We can achieve unitary permeability $\mu'$ by imposing a unitary Jacobian $\det(\bar{\bar{J}})=1$. The geometric interpretation is that the transformation is area-preserving, such that the area of the triangle *AOB* is always equal to that of *A'O'B'*.

Next we consider permittivity. Analytically, the permittivity tensor $\bar{\bar{\varepsilon}}$ can be solved given the position of *AOB* and *A'O'B'*. The solution can be simplified under the assumption that the transformation takes place only along the *x* axis [*i.e.* $y'=y$ as shown in Fig. 1(b)]. In this case the heights of *A'O'B'* and of *AOB* are equal. To guarantee area preservation, we set the bottom length of the two triangles equal, *i.e.* |*A'B'*|=|*AB*|. The transformation function now becomes a horizontal shear: $x'=x+by$, $y'=y$, where $b=d_O/h$, $d_O$ and $h$ are labelled in Fig. 1(b). The relative permittivity and permeability become:

$$\bar{\bar{\varepsilon}}' = \begin{bmatrix} b^2+1 & b \\ b & 1 \end{bmatrix}, \quad \mu'=1. \quad (1)$$

We define the anisotropy ratio as $R \equiv n_2/n_1 = \sqrt{\varepsilon_2/\varepsilon_1}$, where $n_1$ and $n_2$ ($n_1 > n_2$) are the two principal refractive indices along two orthogonal directions, and $\varepsilon_1$ and $\varepsilon_2$ are the corresponding eigenvalues of the permittivity tensor $\bar{\bar{\varepsilon}}'$. *R* can be obtained analytically as

$$R = \left[ \frac{(b^2+2)/2 - \sqrt{(b^2+2)^2/4 - 1}}{(b^2+2)/2 + \sqrt{(b^2+2)^2/4 - 1}} \right]^{\frac{1}{2}}. \quad (2)$$

Although several reports have discussed the practical concerns of achievable values for either permittivity or permeability [21-23], there has been much less discussion on the practically achievable values of *R*.

As a case study of a 2D TM beam bending adapter, we will illustrate the comparison between a boundary-preserving transform (BPT) method [Fig. 2(a)] and the area-preserving transformation (APT) method [Fig. 2(b)]. As shown in both figures, the rectangular region *AOBC* is a part of a rectangular planar waveguide. To form one arm of the bending adapter, we transform *ΔAOB* to the new triangle *ΔAO'B*, where the angle $\alpha$ is the half-bending angle of the adapter. The final bending angle is $2\alpha$, formed by mirroring the structure with regard to axis *BO'*, as shown in the inset figure. Without loss of generality, we set length |*OB*|=1, and fix the point *A* on the horizontal axis at location (-*L*, 0). Besides these common parts, the difference between the two figures is the position of *O'*.

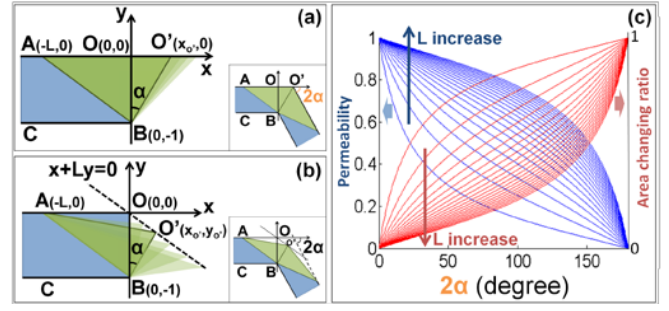

**Figure 2.** (Color online) (a) BPT (from *ABO* to *AB'O'*) for one arm of the bending adapter. (b) APT for one arm of the bending adapter. (c) In BPT case, the permeability (red) and the area changing ratio (blue) as functions of bending angle $2\alpha$ and the arm length *L*.

We first discuss the BPT method in Fig. 2(a), where the location of *O'* is determined from the boundary-preserving requirement, *i.e.* that the geometrical conditions *O'A*//*BC* and ∠*OAB* may not be violated; thus, *O'* should remain on the horizontal axis. This strategy has been common in many earlier designs [8, 17, 24]. Figure 2(c), corresponding to BPT, shows that the permeability indeed varies dramatically with the bending angle $2\alpha$ (blue curves) and cannot equal one except in the trivial case *O*=*O'*. The variance is in accordance with the area changing ratio (red curves) defined as $(S_a - S_b)/S_a$, where $S_a$ and $S_b$ denote the areas after and before transformation, respectively. The area change is directly related to the amount of nonunit permeability $\mu'$. Interestingly, if the permeability is renormalized to unity, the resulting anisotropic medium is equivalent to an isotropic medium with impedance $\eta_0/\mu'$, where $\eta_0$ is the impedance of the original medium before transformation. Therefore $\mu'$ also measures the impedance mismatch in the renormalization.

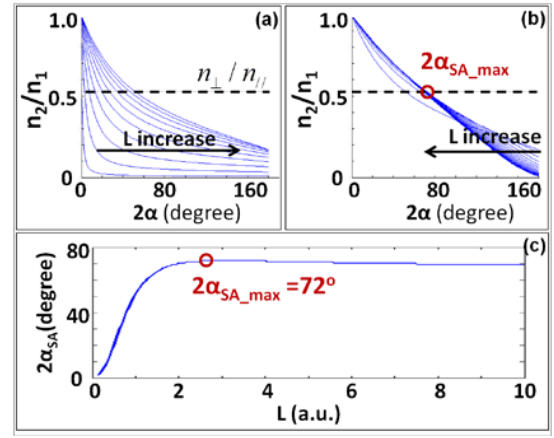

**Figure 3.** (Color online) (a) Anisotropy ratio $R = n_1/n_2$ as the function of bending angle $2\alpha$ and arm length *L* (*L*</*OB*|). The dashed line corresponds to the value of $n_\perp/n_\parallel$ for the silicon-air material system. (b) Same as (a), except |*OB*|<*L*<10|*OB*|. (c) Maximum bending angle $2\alpha_{SA}$ as a function of *L*.

In contrast, the APT design is shown in Fig. 2(b). Here we require that the area of triangle *AOB* should equal that of *AO'B*. Thus, *O'* should be placed such that *OO'* // *AB*, *i.e.* $x_O + Ly_O = 0$. The Jacobian is thus unitary, leading to $\mu'=1$.

It is a special case ($\angle AOB = 90^\circ$) of the horizontal shear APT [Fig. 1(b)].

Next we proceed to consider the practical limit of achievable anisotropy in APT. $R$ depends on the bending angle $2\alpha$ ($0 < 2\alpha < 180^\circ$) given a certain arm length $L$ according to (2), and shown in Fig. 3(a) ($L < |OB|$) and (b) ($L > |OB|$). For fixed arm length $L = |OA|$, increasing $2\alpha$ results in larger shear, and thus smaller $R$. On the other hand, for fixed $2\alpha$, varying $L$ will also results in increased $R$ when $L < |OB|$, but decreased $R$ when $L > |OB|$ for certain value of $2\alpha$. The amount of allowable anisotropy can thus be determined from these plots, taking into the account the available anisotropy of our chosen optical material or nanofabrication method.

As an example, we analyze the largest bending angle that can be achieved using silicon-air layered structure with subwavelength period $\lambda/10$. The refractive indices of silicon and air are $n_{Si} = 3.48$ and $n_{Air} = 1$ respectively at $\lambda = 1550$ nm. From effective medium theory, the parallel and perpendicular effective permittivities are, $\varepsilon_{//} = r\varepsilon_{Si} + (1-r)\varepsilon_{Air}$ and $\varepsilon_\perp = \varepsilon_{Si}\varepsilon_{Air}/[r\varepsilon_{Air} + (1-r)\varepsilon_{Si}]$ respectively. The filling factor $r$ is set as 0.5 to achieve maximum anisotropy, i.e. the smallest ratio of $n_\perp/n_{//} = \sqrt{\varepsilon_\perp/\varepsilon_{//}} = 0.53$, marked as the dashed line crossing Fig. 3(a) and (b). If the required value of $R$ (the anisotropy ratio) is larger than $n_\perp/n_{//}$, it can be achieved by adjusting the filling factor $r$. Otherwise, the silicon-air structure would not possess sufficient anisotropy to realize this bending angle. Setting $n_1/n_2 = n_\perp/n_{//}$ we obtain the largest achievable bending angle $2\alpha_{SA\_MAX} = 72^\circ$. In Fig. 3 (a) and (b), this equivalence is shown as crossing points between the line $n_\perp/n_{//}$ and the curves of $n_2/n_1$ with certain $L$ value. Maximum bending is obtained when $L = 2.53|OB|$, as shown by the rightmost crossing point [Fig. 3(b)].

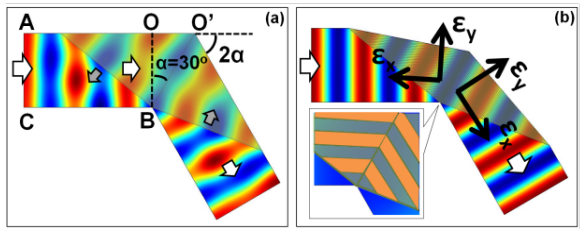

**Figure 4.** (Color online) (a) The magnetic field distribution of the BPT case with scaled $\mu$ and $\varepsilon$. (b) The magnetic field distribution of the APT case. The inset figure demonstrates the Silicon-air layered structure.

To confirm the effectiveness of the wave bending adapter, we performed numerical simulations using the commercial FEM solver, COMSOL Multiphysics. As example, we chose a 60° adapter (i.e., $\alpha$ is 30°). The lengths of $OB$ and $OA$ were set as 1.2 $\mu$m and 1.8 $\mu$m, respectively. Flint glass with refractive index $n_d = 1.87$ was chosen as the waveguide material. A TM planar wave at wavelength 1550 nm is incident from the left waveguide. Figure 4(a) shows the distribution of the magnetic field H in the BPT adapter, with $\mu'$ normalized to unity and $\varepsilon_1$ ($\varepsilon_2$) scaled for nonmagnetism. Figure 4(b) shows H in the APT adapter. The silicon-air interfaces have certain angles (-2.64° for the left arm and -57.36° for the right arm) with respect to the x axis, creating desired permittivity tensor. The transmission in the BPT and APT cases was found to be 92.37% and 100%, respectively. In the BPT adapter it will decrease dramatically if the bending angle is increased further. The energy transmission loss of 60 degree BPT is similar to that through a glass slab. However, the beam profile has been seriously distorted, which will affect the signal delivery in potential optoelectronic application. In contrast, the beam profile is well preserved in the APT adapter.

In conclusion, we have introduced a general nonmagnetic affine transformation method and illustrated a practical design of the 2D TM bending adapter. The beam profile is well preserved with almost ideal transmission in a nonmagnetic realization. Our detailed analysis of the achievable bending angle limited by anisotropy should provide guidance for future practical fabrication in both the macroscopic and nanoscale situations.